# Fractal-driven distortion of resting state functional networks in fMRI: a simulation study

Wonsang You[1], Jörg Stadler[1]
[1] Special Lab Non-invasive Brain Imaging, Leibniz Institute for Neurobiology, Magdeburg, Germany
E-mail: you@lin-magdeburg.de

Fractals are self-similar and scale-invariant patterns found ubiquitously in nature. A lot of evidences implying fractal properties such as 1/f power spectrums have been also observed in resting state fMRI time series. While the traditional model of fractal behavior in resting state fMRI has been a fractional Gaussian noise, it is limited to describe the physical implication of fractal behavior on functional connectivity of the brain. To answer this problem, we have proposed the fractal-based model of resting state hemodynamic response function (rs-HRF) whose properties can be summarized by a fractal exponent (*You et al. 2012 BMC Neurosci.*). Here we show, through a simulation studies, that the fractal behavior of cerebral hemodynamics may cause significant distortion of network properties between neuronal activities and BOLD signals. We simulated neuronal population activities based on the stochastic neural field model from the Macaque brain network, and then obtained their corresponding BOLD signals by convolving them with the rs-HRF filter. The precision of centrality estimated in each node was deteriorated overall in three networks based on transfer entropy, mutual information, and Pearson correlation; particularly the distortion of transfer entropy was more sensitive to the standard deviation of fractal exponents (Figure 1). A node with high centrality was resilient to desynchronized fractal dynamics over all frequencies while a node with small centrality exhibited huge distortion of both wavelet correlation and centrality over low frequencies (Figure 2). This theoretical expectation indicates that the difference of fractal exponents between brain regions leads to discrepancy of statistical network properties, especially at nodes with small centrality, between neuronal activities and BOLD signals, and that the traditional definitions of resting state functional connectivity may not effectively reflect the dynamics of spontaneous neuronal activities. As an alternative, the *nonfractal connectivity*, which is defined as the correlation of nonfractal components of resting state BOLD signals, can be considered to overcome the fractal artifact (*You et al. 2012 IJCNN*). In conclusion, our simulation studies may give us insight into the influence of fractal behavior on complex networks of the brain.

**References**

1. Werner G: **Fractals in nervous system: conceptual implications for theoretical neuroscience**. *Front Physio* 2010, **1**:15.
2. You W, Achard S, Stadler J: **Fractal-based linear model of resting state hemodynamic response in fMRI**. *BMC Neuroscience 2012*.
3. You W, Achard S, Stadler J, Brückner B, Seiffert U: **Fractal analysis of resting state functional connectivity of the brain**. *2012 International Joint Conference on Neural Networks*.

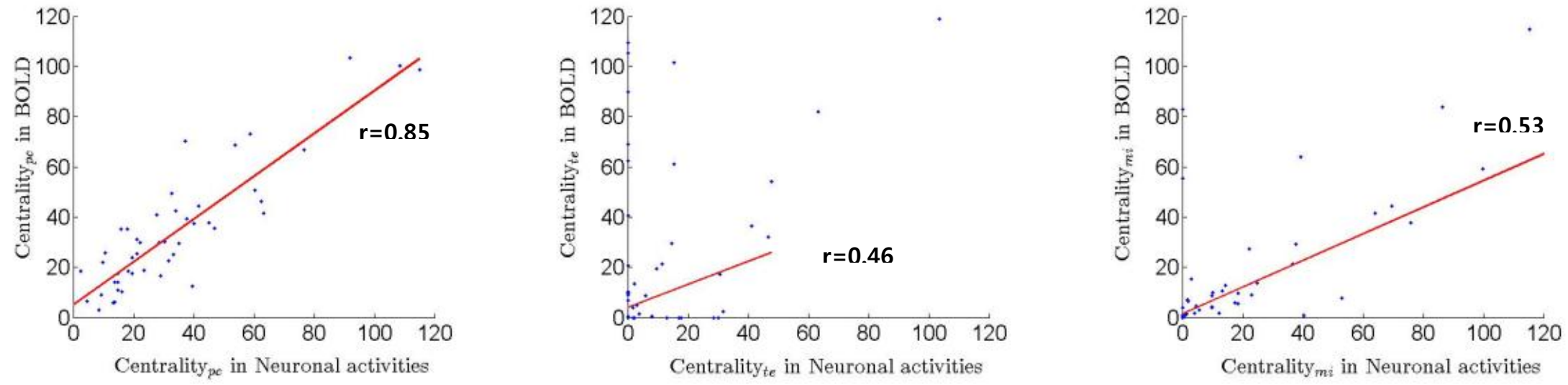

(a) $SD(d)$ = 0.1

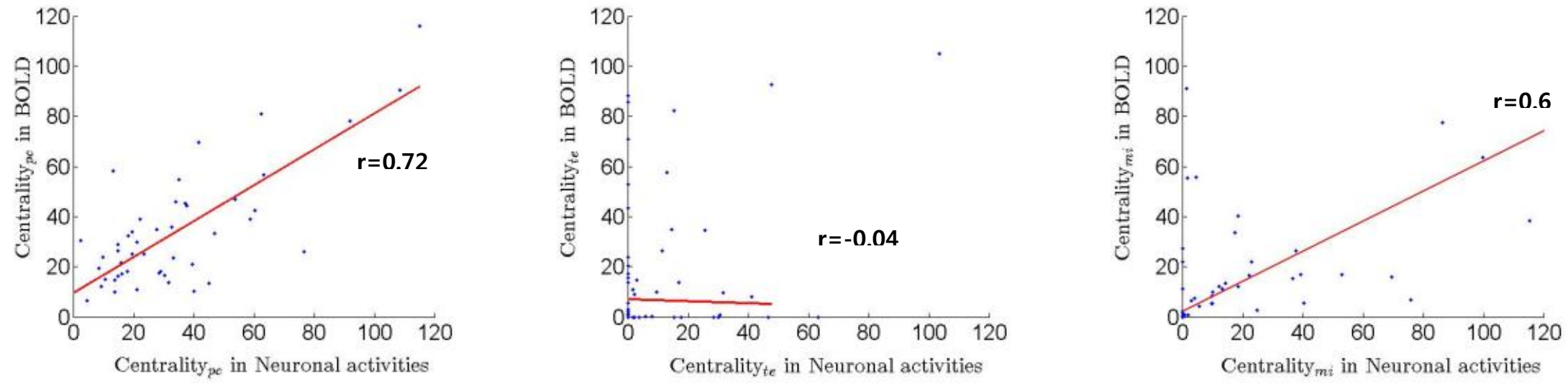

(b) $SD(d)$ = 0.3

**Figure 1**. Scatter plots which illustrate the differences of centrality in three types of functional networks such as Pearson correlation (*left*), transfer entropy (*middle*), mutual information (*right*) between neuronal activities and BOLD signals. Two cases, when the standard deviations of fractal exponents are (a) 0.1 and (b) 0.3, are compared. In Pearson correlation, the regression slope decreased slightly when the standard deviation of fractal exponents was high. The decrease of the regression slope was more abrupt in transfer entropy, and resulted in the increase in difference of neuronal activities and BOLD signals. On the other hand, the slope in mutual information was less affected by the standard deviation of fractal exponents; however the absolute values were smaller than those of Pearson correlation.

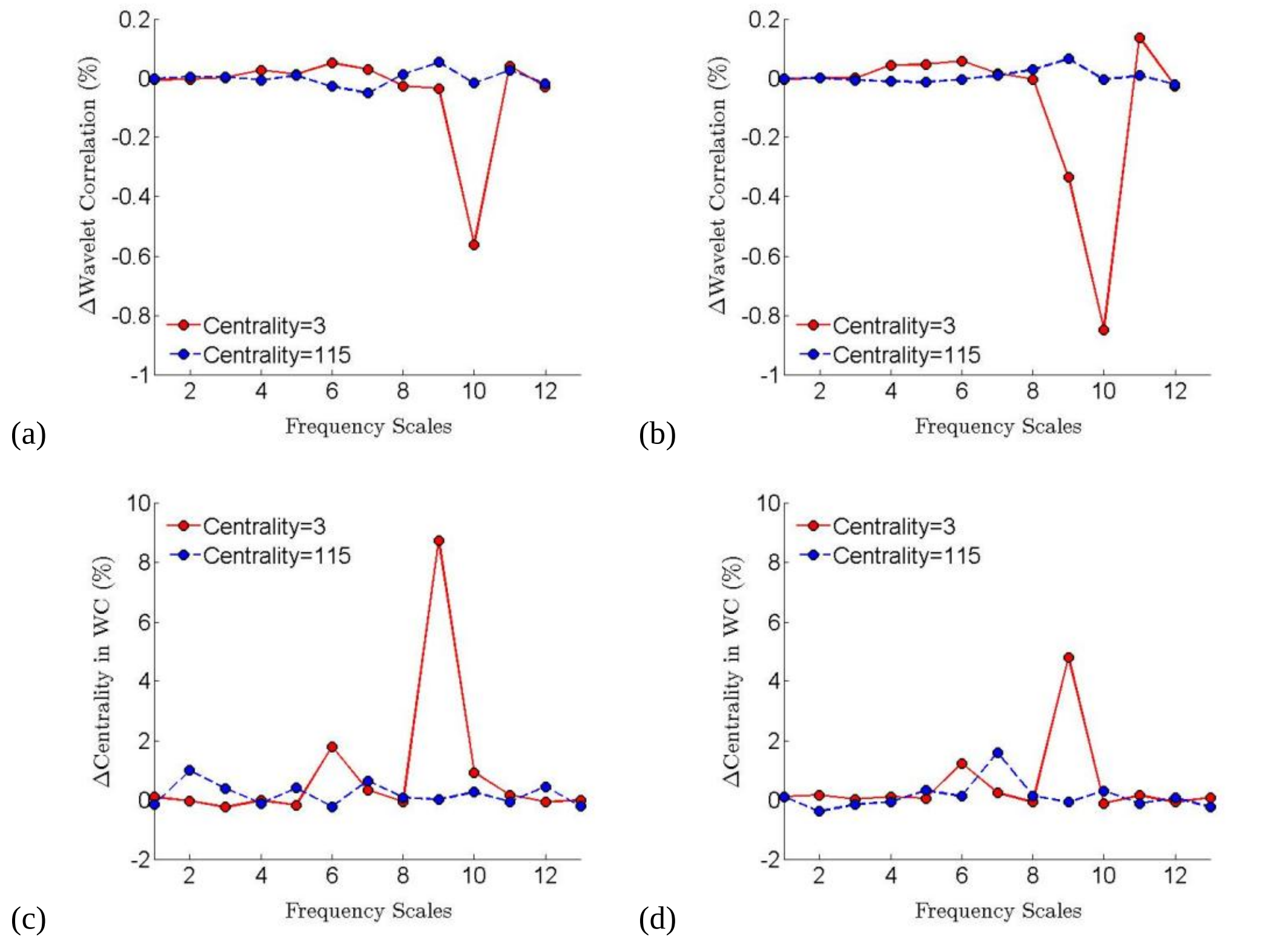

**Figure 2**. (a-b) The relative deviation of wavelet correlations over scales (where high scale corresponds to low frequencies), and (c-d) the deviation of centrality over scales. Left and right are for *SD*(*d*) = 0.1 and 0.3 respectively. For a node with small centrality, as the standard deviation of fractal exponents increases, the absolute peak of deviation in wavelet correlation also increases in low frequency ranges while the deviation in centrality decreases. It is obvious that the dissimilarity of either wavelet correlation or centrality between neuronal activities and BOLD signals becomes larger in a node with small centrality than one with high centrality.